# Synergy of Volunteer Measurements and Volunteer Computing for Effective Data Collecting, Processing, Simulating and Analyzing on a Worldwide Scale

N. Gordienko*, O. Lodygensky**, G. Fedak*** and Yu. Gordienko****
* Phys.-Math. Lyceum 142, Kyiv, Ukraine
** LAL, University Paris South, Orsay, France
*** University of Lyon, Lyon, France
**** G.V.Kurdyumov Institute for Metal Physics, National Academy of Sciences, Kyiv, Ukraine
e-mail: assasin.nik@gmail.com

**Abstract** - The paper concerns the hype idea of "Citizen Science" and the related paradigm shift: to go from the passive "volunteer computing" to other volunteer actions like "volunteer measurements" under guidance of scientists. They can be carried out by ordinary people with standard computing gadgets (smartphone, tablet, etc.) and the various standard sensors in them. Here the special attention is paid to the system of volunteer scientific measurements to study air showers caused by cosmic rays. The technical implementation is based on integration of data about registered night flashes (by radiometric software) in shielded camera chip, synchronized time and GPS-data in ordinary gadgets: to identify night "air showers" of elementary particles; to analyze the frequency and to map the distribution of "air showers" in the densely populated cities. The project currently includes the students of the National Technical University of Ukraine "KPI", which are compactly located in Kyiv city and contribute their volunteer measurements. The technology would be very effective for other applications also, especially if it will be automated (e.g., on the basis of XtremWeb or/and BOINC technologies for distributed computing) and used in some small area with many volunteers, e.g. in local communities (Corporative/Community Crowd Computing).

## I. Introduction

The proposed topic concerns the new hype idea of "Citizen Science" and volunteer involvement in science [1]. The main principle here is the paradigm shift: to go from the passive "volunteer computing" (widely used now in many fields of sciences [2-5]) to other volunteer actions under guidance of scientists as "volunteer measurements", "volunteer data processing and visualization", "volunteer data mining", etc. They can be carried out by ordinary people with modern standard computing units (e.g. wearable PC, smartphone, tablet, smart watch, etc.) with various operational systems (Android, iOS, Windows, Tizen, etc.) and the standard sensors in them or easily accessible external measuring units.

The main components of the idea or conditions for effective implementation of volunteer measurements are as follows: (a) they should be used by ordinary people – not scientists only, (b) they should be performed by easily accessible measuring units – like modern smartphones with numerous sensors, and (c) in unobtrusive way.

The most important expectations (and potential criteria of success) are possibilities to:

- leverage the available "crowdsource" resources: machine (personal CPUs + sensors) and human (brains, manual operations of sensors) ones,

- get huge number of volunteers (up to millions),

- obtain the new scientific "quality" from these huge "quantities",

- involve ordinary citizens in scientific process,

- report to society about the current scientific activities and priorities.

## II. Previous Work and State of the Art

The feasibility of the proposed idea was already tested in several awarded volunteer projects with the several types of activities in different fields of science, which is illustrated by some use cases below in this section.

### A. Monitoring Environmental Conditions

In 2012 "My Green City" project was initiated by the group of students for monitoring environmental conditions and their effects on everyday life [6]. It was motivated by the famous problem that vehicle exhausts are ones of the most dangerous sources of air pollution in cities, but global attempts to assess how emissions impact local city ecology have proven to be ineffective.

The aims of "My Green City" project were as follows:

- to estimate personal dosage of accumulated dangerous emissions by volunteers,

The work presented here was partially supported by EU FP7 SCI-BUS (SCIentific gateway Based User Support) project, No. RI-283481.

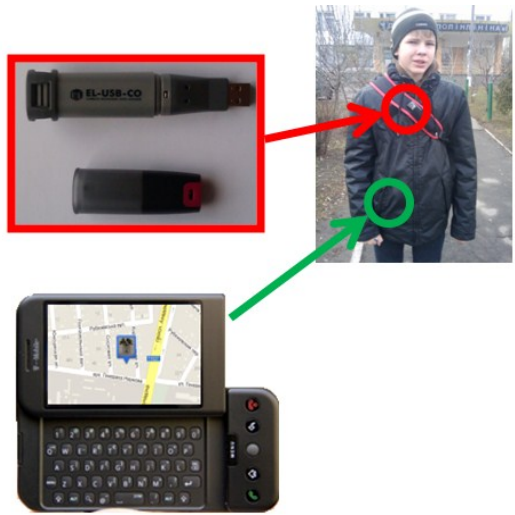

Figure 1. The typical volunteer gear: CO-gas-USB logger (top left), which can be attached to rucksack, and smartphone with GPS-navigator in the pocket (bottom left) of the volunteer (right)

- to determine real-time distribution of pollutants by vehicle exhausts in cities.

The technical implementation was based on integration of pollution data by vehicle emissions sensor (CO-gas-USB logger) and location data by GPS-navigator in ordinary smartphones (Fig. 1) for: estimation of personal dosage of accumulated dangerous emissions (so called "personal air health monitor"), and continuous air pollution data processing and construction of interactive air pollution map (Fig. 2).

To check feasibility of this proposal the student pilot project "My Green City" was implemented with construction of the air pollution map for the city of Kyiv (Ukraine) in 2012. The social incentive behind this pilot project can briefly be expressed by the slogan: "from volunteer scientific calculations to volunteer scientific measurements". It became one of the 15 global finalists in Google Science Fair 2012 [7].

Scientists can develop tools (personal air health monitor and online virtual air pollution map) by integration of location-based data from sensors of numerous volunteers. It can allow volunteers and communities to make real time estimations about the actual impact of vehicle emissions on a city environment.

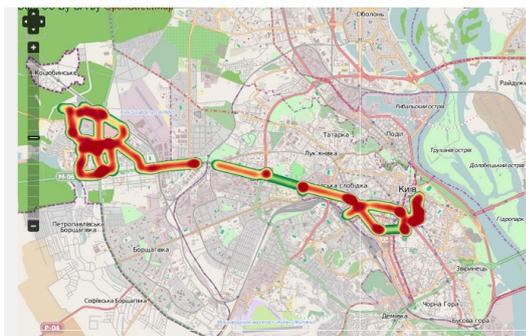

Figure 3. The interactive air pollution map for the city of Kyiv along the regular tracks of children acting as volunteers
(http://dg.imp.kiev.ua/slinca/my-green-city/CO)

For municipal authorities, these tools could improve overall city traffic planning and city management accountability. For people, these tools could stimulate better personal planning and implement healthier behavior due to: (A) the greater public awareness of air pollution distribution by virtual air pollution map; and (B) the targeted estimation of personal dosage of accumulated dangerous emissions by personal air health monitor.

*B. Monitoring Personal Health*

The other volunteer measurement project «How to Make Old Age More Comfortable, Less Boring and Active for Your Nearest and Dearest» (2013) was dedicated to the following problem [8]. The human population become older and will continue to do so in the coming decades; elderly people suffer from various diseases. The project activities were based on the fact that elderly people suffer from various diseases, but their progress can be monitored by decrease of mobility. The aim of the project was to explore the use of sensors in the usual smartphone to monitor the health of older people by analysis of their locomotor activity. The similar method was used in this project, namely, embedded personal sensors (accelerometers) and smart phones were used for real time monitoring the motion pattern (mean, standard deviation, skewness, curtosis, etc.) for different types of motion (sleep, walk, run, etc.).

The technical implementation (Fig. 3) consisted in integration of personal measurement tools and attachment points on the basis of the following available gadgets:

- smartphone with accelerometer (G-sensor) on the arms (during writing, web-surfing, etc.) and legs during cycling (walking, running, etc.);
- "smart watch" (the arm wrist device ez430 by Texas Instruments) with accelerometer and wireless data broadcasting to a PC.

Multi-parametric moment analysis was applied for data analysis. And the previous results shown that activities can be roughly classified, i.e. divided into groups (colored ellipses) with the similar values of the acceleration distribution parameters (Fig. 4, see legend in the color electronic version):

- passive (web surfing, read, sleep) (brown ellipse),
- moderate (writing, sitting) (green ellipse),
- active (sports, housework, walking) (blue ellipse).

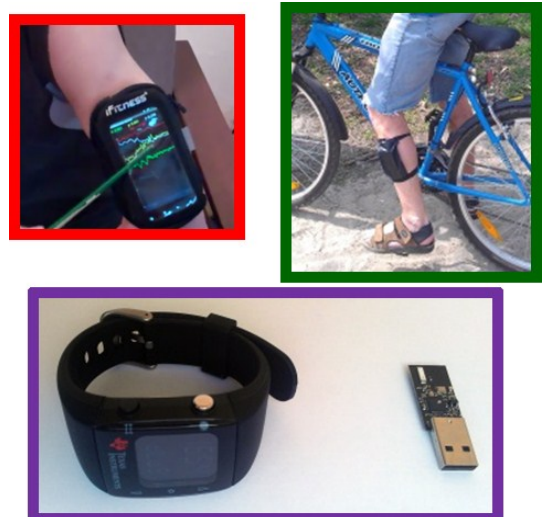

Figure 2. The typical location of measurement gear on volunteers: HTC smartphone with G-sensor on the arm during writing, web-surfing, etc. (top left) and on the leg during biking (top right), and "smart watch" (ez430, Texas Instruments) (bottom)

To check feasibility of this idea the student project "How to Make Old Age More Comfortable, Less Boring and Active for Your Nearest and Dearest" (2013) was implemented in the local community. Again, the social incentive behind this pilot project can be expressed by the slightly different slogan: "from general to personal volunteer scientific measurements". It became one of the 90 Regional Finalists in Google Science Fair 2013 [9].

The tests have shown that some essential features of health and activity patterns can be distinguished and classified. By analysis of sensor (accelerometer) data scientists can develop tools (so called "personal activity monitor") that allow volunteers and communities to make real time estimations of the actual locomotor activity decrease and subsequent health decay. For health care authorities, these tools could improve overall elderly care and stimulate personal online elderly care services. For communities, these tools could stimulate closer attention to locomotor activity and its automated monitoring due to: (A) the greater public awareness of dangerous locomotor activity decrease and probable subsequent health decay; (B) the targeted estimation of personal dangerous decrease of locomotor activity by personal activity monitor; (C) increase of efficiency of elderly care in some communities by means of automatic detection of dangerous decrease of locomotor activities among neighbors, (D) decrease of burden on communities, which is related with a high costs of elderly care.

### III. CURRENT PROJECT - AIRSHOWER@HOME

#### A. Scientific Problem

Cosmic rays originate mainly outside the Solar System. They may produce showers of secondary particles ("air showers") that penetrate and impact the Earth's atmosphere and sometimes even reach the surface. "Air showers" of secondary elementary particles affect electronics, health, environment. Their study is expensive (>$50 mln.) and limited in scale. The most energetic cosmic rays (>$5 \times 10^{19}$ eV) are most important, but very rare were observed, for example, 1 high-energy event is registered per month by Pierre Auger observatory. Pierre Auger Observatory — is an international cosmic ray observatory designed to detect ultra-high-energy cosmic rays [10,11]. It detects cosmic rays in the range of $10^{17}$-$10^{21}$ eV, costs >$50 mln., and has a detection area >12 km$^2$ and huge installation area >3000 km$^2$ (like Luxemburg). Cosmic rays affect electronics, because they have sufficient energy to alter the states of circuit components in electronic integrated circuits, causing transient errors like corrupted data in electronic memory devices, and incorrect performance of CPUs, often referred to as "soft errors" (not to be confused with software errors caused by programming mistakes/bugs!). They affect health, especially the health threat from cosmic rays is the danger posed by galactic cosmic rays and solar energetic particles to astronauts on interplanetary missions. They are one of the most important barriers standing in the way of plans for interplanetary travel by crewed spacecraft [12,13]. Cosmic rays affect environment directly or via solar-induced modulations in climate change, because solar variations modulate the cosmic ray flux on Earth, they would consequently affect the rate of cloud formation and hence the climate [14].

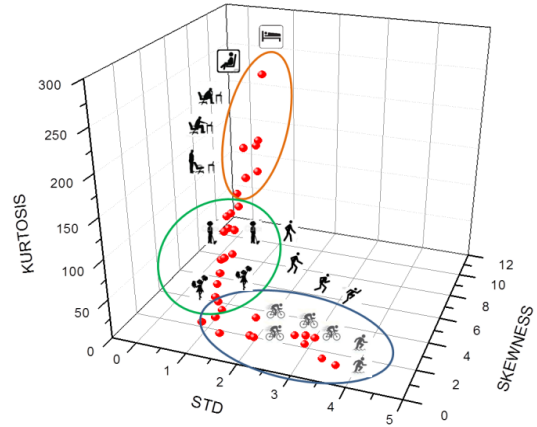

Figure 4. The moments (standard deviation, skewness, kurtosis) plot: activities (denoted by pictograms) can be classified by groups (colored ellipses in the color electronic version) with the similar values of the acceleration distribution parameters: passive (brown ellipse), moderate (green ellipse), active (blue ellipse)

#### B. AirShower@Home Project Description

As far as "air showers" of elementary particles created by cosmic rays affect electronics, they can be monitored by some available devices (like CMOS-camera chips).

The aims of AirShower@Home project were:

- to estimate possibility to identify "air showers" of secondary elementary particles,
- to determine frequency and distribution of "air showers" in the densely populated cities – by massive involvement of volunteers with creation of virtual online map of "air showers".

The volunteer measurements could be performed on the basis of the following tools:

- camera chips as embedded sensors in GPS-enabled smartphones;
- volunteer communities in densely populated locations for real time monitoring the "air showers" and analysis of their effects (on environment, electronics, health).

The technical implementation is based on integration of data about registered flashes (by radiometric software) in shielded camera chip, synchronized time and GPS-data in ordinary smartphones/tablets/other gadgets. The current workflow is shown in Fig. 5, where the size of the available academic solution, Pierre Auger observatory, is

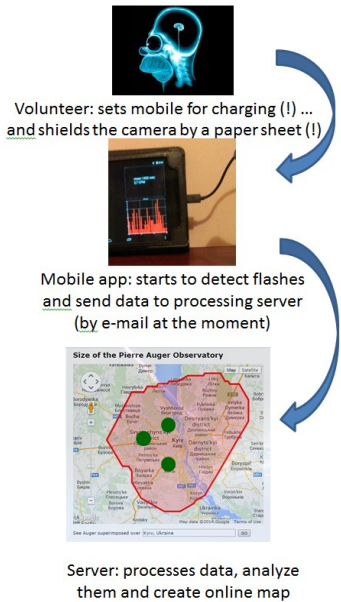

Figure 5. The diagram of the current workflow, where the sizes of Pierre Auger observatory are shown for comparison on the map of measurements in the city of Kyiv (bottom)

shown for comparison on the map of measurements in Kyiv. It includes the following components:

- hardware (various gadgets were tested: Samsung SIII (with 8 megapixels camera), HTC Amaze smartphone (8 Mpx), Nexus 7 tablet (1.2 Mpx), Nexus 7 tablet (5 Mpx), ASUS TF201 tablet (8 Mpx),
- "brainware", because some human actions are needed (volunteer should shield the camera, and perform manual post-processing, analyzing and mapping "air showers"),
- software (various mobile apps): radiation counters (used for tests only), apps for time synchronization, and e-mail clients for submissions of logs,
- volunteer community: >80 pupils and students were involved already (during the lab works in "Distributed Computing Course" at the National Technical University of Ukraine "KPI" in Kyiv).

IV. RESULTS OF THE CURRENT PROJECT

One of the examples of the registered radiation results in counts per minute ("cpm") vs. time (within ~5 days) is shown in Fig. 6. The four smooth peaks corresponds to slow increases of background radiation (caused by the temperature increase during days), and numerous sharp peaks potentially correspond to cases of air showers caused by cosmic rays. The different colors indicate different values of filters used in the measurements. These results lead to conclusion that seasonal and other temperature changes could crucially influence the measured values of radiation and should be taken into account during such crowd measuring/computing projects.

The other example is related to tests performed on a board of the commercial air flight. The idea was to check feasibility of volunteer measurements of air showers on various heights using the standard commercial flights (with the usual safety precautions as to usage of electronics during commercial flights). Registration of air showers was performed simultaneously by two Nexus tablets (with different sensitivity) on a board of the Airbus 320 plane during flight Kyiv-Budapest on 10/03/2014. The radiation (in cpm) as a function of time was measured (Fig. 7), where smooth growth corresponds to increase of the registered radiation counts with altitude (9 km), and the same sharp peaks correspond to the events of simultaneous registration of the air showers caused by cosmic rays. As to possible pitfalls in such experiments, the problems related with gadget overheating can be also observed in the end of flight.

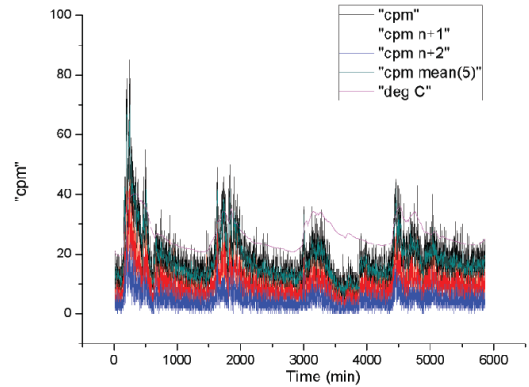

Figure 6. Registered radiation (counts per minute — cpm) vs. time

The initial stage of this crowd measuring/computing idea was performed among numerous volunteers from Lyceum 142 and the National Technical University of Ukraine "KPI" (Fig. 8). Mainly they were compactly located in the city of Kyiv (Fig. 8, inset), and this local concentration of volunteers is especially useful for the current scientific application. The measured data were send by mobile application to server, where they were collected, post-processed (in R environment for statistical analysis), mapped (by means of googleVis and Shiny packages for R environment), and published on the AirShower@Home project web-page at ShinyApps.io server (http://yocto.shinyapps.io/Height_test).

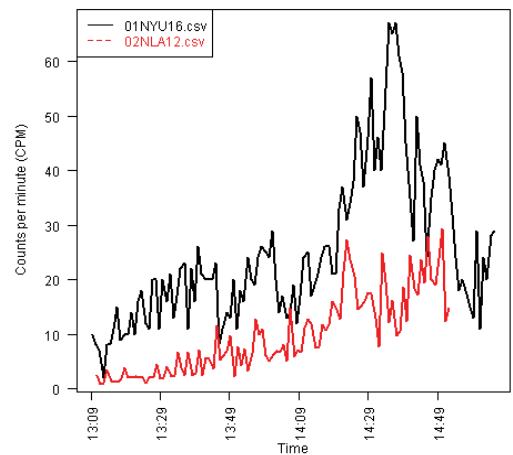

Figure 7. Registration of air showers during a commercial flight

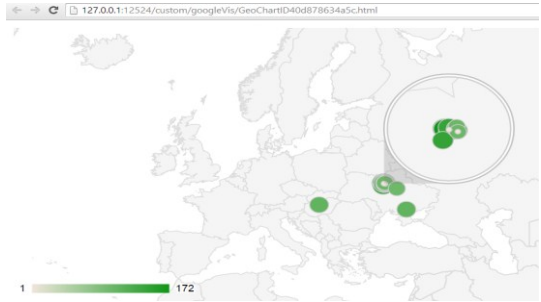

Figure 8. The map of heights at volunteers' locations among other maps with post-processed data of AirShower@Home project (http://yocto.shinyapps.io/Height_test)

Actually, these interactive maps contain information about volunteer locations, date and time of volunteer measurements, height of their locations (which are important for comparison), and the registered radiation counts.

In Table 1 the short comparative analysis is given for several solutions, namely, for academic solution (Pierre Auger observatory), do-it-yourself (DIY) solution with external sensor (customized solution on the basis of Raspberry PI single-board computer), and the current AirShower@Home solution with embedded sensor (widely available solution like smartphone). From this analysis it is clear that the AirShower@Home project provides an additional (not alternative!) way to study cosmic ray (by analysis of air showers) and their influence on civilization without huge installations, numerous high-qualified scientists, high basic and maintenance costs. Moreover, its crowd measuring/computing infrastructure can be effectively used for many other applications, for example, in two previous projects to upgrade and scale them up to worldwide scale.

TABLE I. COMPARISON WITH OTHER SOLUTIONS

|   | *Academic solution (Pierre Auger observatory* | *DIY solution with external sensor (Raspberry PI + sensor)* | *This "Citizen Science" solution (smartphone with sensors)* |
|---|---|---|---|
| Size | 3000 km$^2$ | 1 cm$^2$ | 4 mm$^2$ |
| Scalability | <30% | <10 times | > 10$^6$ (unlimited) |
| Maximal Area | ~3000 km$^2$ | ~1 cm$^2$ | >10$^6$ km$^2$ |
| Staff (per installation) | 500 scientists | 1-3 scientists | 1 ordinary human ("Citizen Scientist") |
| Cost | > $50 000 000 | ~$500 | <$200 (smartphone) |

## V. CONCLUSIONS

Some lessons were learned during these three different volunteer measuring/computing projects. First of all, the positive aspects consist in feasibility:

- to leverage the available "crowdsource" resources (hardware and "brain-ware") even at this level – it is feasible, if sensors are embedded already (integration with external sensors is hardly can be attractive for potential volunteers),

- to get huge number of volunteers (millions) – it is feasible for project without sensitive data, but it could be hardly feasible for projects related with personal location and health data due to privacy issues,

- to obtain the new scientific "quality" from these huge "quantity" of data – it is feasible, and porting to the standard platforms for volunteer computing should help, because the limited automation is used at the moment,

- to involve ordinary citizens in scientific process – it is feasible, if task, results, and volunteer management will be fully automated, for example, by porting to BOINC [15] and XtremWeb environments [16],

- to report to society about the current scientific activities and priorities – it is feasible, again after porting to BOINC [15] and XtremWeb environments [16].

Taking into account these considerations and comparative analysis (Table 1) the immediate ways for improvement are seem to be based on porting this crowd measuring/computing infrastructure to the standard platforms for volunteer computing. Some of them (like BOINC and XtremWeb) are proved to be effective for distributed computing applications. For example, XtremWeb was successfully used for AIRES application used in relation to results obtained at Pierre Auger observatory [17]. Additional potential for improvement is related with optimization of the current workflow, for example, on the basis of the Science Gateway ideology. Now it is very promising way for integration and management of tasks, users, and resources [18,19], for example, on the basis of gUSE-WS-PGRADE package [20].

Such porting will allow better integration with embedded sensors and automatic cooperation with them (without manual operations), post-processing, analyzing and mapping results, and effective volunteer community management (job submission, data storage and analysis, forum, etc.) at server, especially, in open local and corporative communities (like schools, colleges, universities). As to potential contribution to science, many scientific phenomena can be investigated by ordinary people and in any place, not just in some secluded labs, which is essential for "Citizen Science" paradigm. This approach could be very useful, if it will be used in some areas with many volunteers, e.g. in local communities (which could be considered as Corporative/Community Crowd Computing)

ACKNOWLEDGMENT

The authors would like to thank numerous volunteers from Lyceum 142 and the National Technical University of Ukraine "KPI" who took part in the volunteer measurements during the initial stage of the project.

## REFERENCES

[1] A.Irwin, Citizen science: a study of people, expertise, and sustainable development, Psychology Press, 1995.

[2] D.P.Anderson, G.Fedak, "The computational and storage potential of volunteer computing", In. Proc. Sixth IEEE International Symposium on Cluster Computing and the Grid (CCGRID 06), Vol. 1, pp. 73-80, May, 2006.

[3] O.Baskova, O.Gatsenko, Yu.G.Gordienko, "Enabling High-Performance Distributed Computing to e-Science by Integration of 4th Generation Language Environments with Desktop Grid Architecture and Convergence with Global Computing Grid", Proc. Cracow Grid Workshop'10, Cracow, Poland, pp. 234-243 (http://www.researchgate.net/publication/261361640) 2010.

[4] O.Gatsenko, O.Baskova, O.Lodygensky, G. Fedak, Y.Gordienko, "Statistical properties of deformed single-crystal surface under real-time video monitoring and processing in the desktop grid distributed computing environment", Key Engineering Materials, 465, pp. 306-309, 2011.

[5] O.Gatsenko, L.Bekenev, E.Pavlov, Yu.Gordienko, "From Quantity to Quality: Massive Molecular Dynamics Simulation Of Nanostructures under Plastic Deformation in Desktop and Service Grid Distributed Computing Infrastructure", Computer Science, Vol. 14, 1, pp. 27-44, 2013.

[6] M.Klimenko, O.Kozlov, "My Green City" project, Google Science Fair 2012 (http://goo.gl/o2qmL1).

[7] "Google Announces 15 Finalists for 2012 Google Science Fair", BuzzIT: Buzz in Technology (http://goo.gl/3PcbW7) June 6, 2012.

[8] N.Gordienko, "How to Make Old Age More Comfortable, Less Boring and Active for Your Nearest and Dearest", Google Science Fair 2013 (http://goo.gl/2Tl200).

[9] J.Weinberg, "Google Science Fair 2013: Banana breast implants and squid submarines - the best ideas from teen science competition", Mirror (http://goo.gl/WzJdFt) 26 September 2013.

[10] P.Auger, P.Ehrenfest, R.Maze, J.Daudin, R.A.Fréon, Extensive Cosmic-Ray Showers, Rev. Mod. Phys. 11 (3–4), pp. 288–291, 1939.

[11] J.W.Cronin, T.K.Gaisser, and S.P.Swordy, Cosmic Rays at the Energy Frontier, Scientific American, pp. 44 (1998).

[12] Biomedical Results From Apollo - Radiation Protection and Instrumentation (http://lsda.jsc.nasa.gov/books/apollo/S2ch3.htm).

[13] Evaluation of the Cosmic Ray Exposure of Aircraft Crew (http://cordis.europa.eu/documents/documentlibrary/75331981EN6.pdf).

[14] Yu.V.Balabin, A.V.Germanenko, B.B.Gvozdevsky, and E.V.Vashenyuk, "Variations of gamma radiation spectra during precipitations", Journal of Physics: Conference Series, Vol.409, 1, 012243, pp. 1-4 (2013).

[15] D.P.Anderson, "BOINC: A system for public-resource computing and storage", In Proc. Fifth IEEE/ACM International Workshop on Grid Computing, pp. 4-10 (2004).

[16] F.Cappello, S.Djilali, G.Fedak, T.Herault, F.Magniette, V.Néri, O.Lodygensky, "Computing on large-scale distributed systems: XtremWeb architecture, programming models, security, tests and convergence with grid", Future Generation Computer Systems, 21(3), 417-437, 2005.

[17] O.Lodygensky, G.Fedak, V.Neri, A.Cordier, F.Cappello, "Auger & XtremWeb: Monte Carlo computation on a global computing platform", In Proc. Computing in High Energy and Nuclear Physics (CHEP2003) Vol. 40 (2003).

[18] Yu.Gordienko, L.Bekenov, O.Gatsenko, E.Zasimchuk, V.Tatarenko, "Complex Workflow Management and Integration of Distributed Computing Resources by Science Gateway Portal for Molecular Dynamics Simulations in Materials Science", In Proc. Third International Conference "High Performance Computing" HPC-UA 2013 (Ukraine, Kyiv, October 7-11, 2013), pp. 148-155.

[19] Yu.Gordienko, L.Bekenev, O.Baskova, O.Gatsenko, E.Zasimchuk, S.Stirenko, "IMP Science Gateway: from the Portal to the Hub of Virtual Experimental Labs in Materials Science", In Proc. 6th International Workshop on Science Gateways, IEEE Xplore Digital Library, DOI: 10.1109/IWSG.2014.17, pp. 61-66, 2014.

[20] P.Kacsuk, Z.Farkas, M.Kozlovszky, G.Hermann, A.Balasko, K.Karoczkai, I.Marton, "WS-PGRADE/gUSE generic DCI Gateway Framework for a Large Variety of User Communities", Journal of Grid Computing, 10(4), 601-630, 2012.